\documentstyle[preprint,epsfig,prb,aps]{revtex}
\begin{document}
\title{Magnetic hysteresis in the Cu-Al-Mn intermetallic
alloy: experiments and modeling}
\author{Eduard Obrad\'{o}, Eduard Vives, and Antoni Planes}
\address{Departament   d'Estructura i  Constiuents  de la Mat\`eria.\\
Facultat   de   F\'\i sica.  Universitat    de  Barcelona. \\ Diagonal
647. 08028 Barcelona. Catalonia. Spain.}
\tightenlines
\maketitle
\begin{abstract}
We    study  isothermal  magnetization   processes  in  the   Cu-Al-Mn
intermetallic alloy. Hysteresis is observed  at temperatures below the
spin-freezing of  the system. The   characteristics of the  hysteresis
cycles as a function of  temperature and Mn content (magnetic element)
are obtained. At low temperature ($\lesssim 5$ K) a change from smooth
to sharp  cycles  is observed with increasing    Mn content, which  is
related to the decrease of configurational disorder.

We also study  a zero-temperature  site-diluted Ising model,  suitable
for the description of this Cu-Al-Mn  system. The model reproduces the
main   features of the  hysteresis  loops  observed experimentally. It
exhibits  a  disorder-induced  critical  line  separating a disordered
phase from an  incipient  ferromagnetic ground-state.  The  comparison
between the  model and  the  experiments allows  to conclude  that the
observed change in the experimental hysteresis loops can be understood
within the framework of the theory  of disorder-induced criticality in
fluctuationless first-order phase transitions.

\end{abstract}
\pacs{PACS number: }
\section{INTRODUCTION}

Hysteresis is the  history dependence of the  properties of a material
when driven   by an applied external  field  \cite{Hysteresis}.  It is
observed in many different systems in  non-equilibrium conditions.  In
general  it  is  a  dynamical effect  which  must be  treated within a
time-dependent formalism. Here  we will  focus  on a special  class of
systems  that display, under    certain conditions,  time  independent
hysteresis  properties. The   prototypes   are ferromagnets,   at  low
temperature, in which  magnetization  versus magnetic field  paths are
not influenced by  the  applied field rate,  at  least for low  enough
rates. In such  materials, when these paths are  studied in detail, it
is commonly observed that they are not  continuous but rather composed
of small bursts or avalanches:  this is the so-called Barkhausen noise
\cite{Barkhausen19}.  This abrupt    and stochastic  response  of  the
magnetization   to  a  continuously varying external    field has been
related to the disorder existing in the system \cite{Vergne81,Cote91}.
Actually,  domain  walls are  pinned by   impurities  and therefore  a
magnetization change  can  only occur if the   corresponding energetic
barriers are overcome  (depinning). Since  these barriers are   large,
thermal activation is not  effective, and the  evolution of the  sytem
can only proceed by modifying  the external field (athermal behavior).
In  addition to ferromagnetic materials,   a wide  variety of  complex
systems   like martensitic materials\cite{Vives94},    superconducting
films\cite{WuAdams},  or capillary condensation systems\cite{Lilly96},
display similar phenomenology.  All  these systems can be described in
the framework of fluctuationless first-order (FLFO) phase transitions,
i.e.   they exhibit, when driven by  an  external field, a first-order
phase transition   influenced  by  disorder  with  irrelevant  thermal
fluctuations.

Recently, different versions of spin  lattice models with disorder, as
the  Random Bond  \cite{Bertotti90,Vives94b,Vives95} and  Random Field
Ising  models \cite{Sethna93}, have been  used to model hysteresis and
avalanches     in systems    undergoing      FLFO  phase  transitions.
Deterministic algorithms  with  synchronous  dynamics corresponding to
local  energy relaxation have been employed  to study hysteresis paths
at zero  temperature.  It has been found  that the shape of hysteresis
loops depends on the amount of disorder \cite{Vives94b,Sethna93}.  For
low disorder, when sweeping  the field from -$\infty  $ to +$\infty $,
the system reverses sharply the  magnetization (infinite avalanche) at
a given field, while   for  a high amount   of  disorder the path   is
smoother  and extends over a  wide range of fields.  Interestingly the
change  in  the hysteresis  cycle shape  has  been  attributed  to the
existence of a critical point at a given degree of disorder, where the
infinite  avalanche  disappears.   For   such a  degree  of  disorder,
avalanches   of  all sizes are  detected    indicating that the system
behaves critically:  the  avalanche size distribution  follows a power
law characterized by a universal  exponent depending only on the space
dimensionality  \cite{Vives95}.   Actually,   in  agreement  with  the
models,  it has been   experimentally  observed that the amplitude  of
Barkhausen pulses in ferromagnetic materials, and the amplitude of the
acoustic signals generated during martensitic transitions, extend over
several decades indicating  the  possibility of a   universal critical
behavior  \cite{Cote91,Vives94,Babcock90,Carrillo98}.    A  number  of
attempts to make a comparison between models and experiments have been
reported.  \cite{Kuskauer96,Cizeau97}  Nevertheless, the  existence of
criticality  induced by changing the amount   of disorder has not been
confirmed.  The difficulty  arises from the possibility of controlling
the  amount of disorder  in a given system.  In  any case,  it must be
mentioned that the studied models appear  as very simple idealizations
when compared to the complexity inherent to real systems.

This paper is  concerned with the  study of the magnetic properties of
the  Cu-Al-Mn  intermetallic alloy.  After  a  suitable heat treatment
\cite{Obrado98}, this material  displays  an L2$_{1}$ structure.   The
stoichiometric compound  Cu$_{2}$AlMn  is  ferromagnetic  with a Curie
temperature of 630 K and  it is known that  the entire magnetic moment
of  the system must  be attributed  to  the Mn  atoms \cite{Webster88}
coupled via the  RKKY  interaction \cite{Tajima77}.  AC susceptibility
measurements have revealed the existence of a spin-freezing process at
low temperatures.  This  freezing has been  shown  to correspond to  a
spin    glass  phase   transition   for    low    enough Mn    content
\cite{Obrado98b}. Beyond  this problem, here we focus  on the study of
the low temperature magnetic hysteresis for such systems.

We have studied a family  of composition related alloys which undergo,
prior  to the    spin-freezing process,  a  diffusionless   structural
transition to a close-packed phase (martensitic transition).  In these
materials,  magnetic  disorder has  two  main   origins: the lack   of
stoichiometry and the existence of incomplete configurational ordering
\cite{Nakanishi93,Prado98}.  The  important   fact   is that the    Mn
configurational disorder increases  as the Mn  content decreases.   We
show that increasing  the  degree  of  disorder, the  low  temperature
hysteresis  cycle  changes from   sharp  to  smooth.  This  change  is
suggested to be related to a disorder induced phase transition.  Since
the   experimental evidences   are   not  definitively  conclusive, we
reinforce our  interpretation  by developing  a microscopic  model for
such site-diluted magnetic  systems.  By fitting the model  parameters
to Cu-Al-Mn available  data,  we reproduce approximately the  observed
phenomenon  and associate it     with   the disorder  induced    phase
transition.  The  paper is  organized  as follows.  In section  II, we
present the experimental results; in section III  we introduce a model
based on   a site  diluted  Ising model  suitable  for   the study  of
hysteresis  in spin  glass systems and   compare the results  obtained
numerically with  the  experimental ones  corresponding to   Cu-Al-Mn.
Finally, in section IV we discuss the results and conclude.

\section{EXPERIMENTAL RESULTS}

\begin{figure}
\epsfig{file=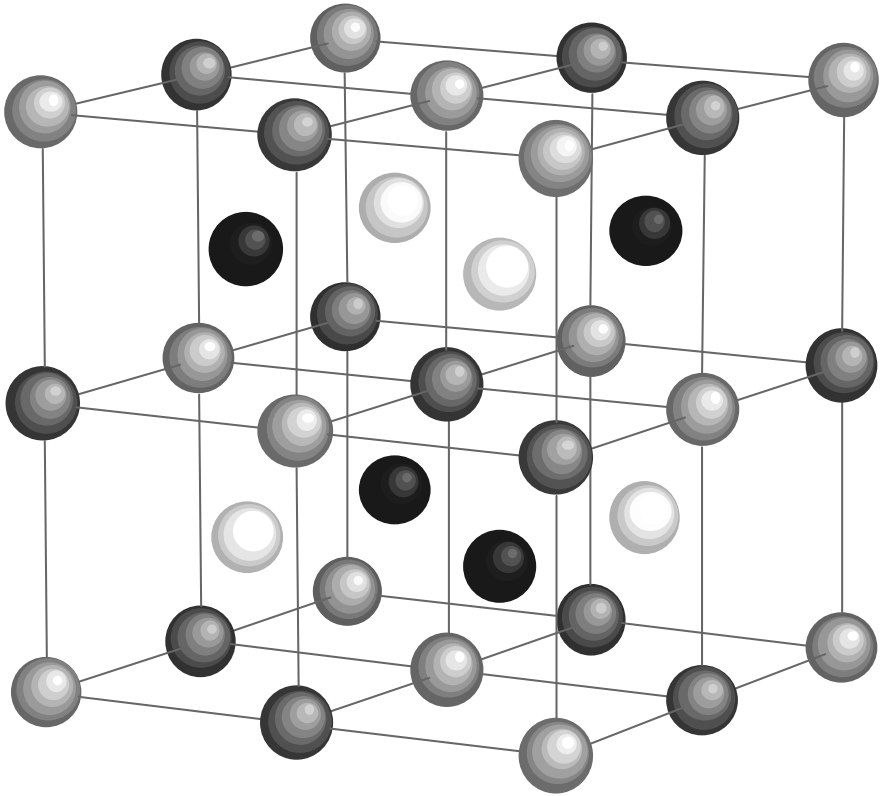,width=12cm}
\caption{Crystal structure  of Cu$_2$AlMn.   Black spheres:  $\alpha $
(Mn)   sublattice.   White spheres:  $\beta   $ (Al)  sublattice. Grey
spheres: $\gamma $ and $\delta $ (Cu) sublattices.}
\label{FIG1}
\end{figure}

We have studied a  family of alloys,  with  compositions close to  the
Cu$_{3-x}$AlMn$_{x}$ line.   Polycrystalline  (grain  size $\sim$  100
$\mu$m)  cylindrical-shaped ingots    were obtained  by   melting pure
elements $(99.99\%)$.    Rectangular shaped  samples  ($9 \times  3.5$
mm$^2$) were   cut using a  low-speed diamond   saw.  The samples were
mechanically polished  down to a thickness  of $0.1$ mm.  Samples were
annealed for $900$ s  at $1073$ K, quenched  into a mixture of ice and
water, and annealed at room temperature  for several weeks.  This heat
treatment leads to a long-lived metastable state in which the material
displays an L2$_{1}$  structure ($Fm3m$ space  group).  This structure
can be  viewed  as  defined  on  a    bcc  lattice  divided  in   four
distinguishible   fcc sublattices  ($\alpha$,   $\beta$,  $\gamma$ and
$\delta$)  as shown in Figure  \ref{FIG1}.  For samples with $x<1$ (as
in the present work), sublattices $\gamma$ and $\delta$ (for instance)
are  filled  with  Cu-atoms, and     Mn-atoms sit  preferentially   in
sublattice $\alpha$.  However, for kinetic reasons,   a fraction of Mn
atoms also occupies the $\beta$ sublattice.  Experiments based on Atom
Location by Channelling Enhanced  Microanalysis (ALCHEMI) performed on
similar samples have shown that almost no  Mn atoms occupy sublattices
$\gamma$ and $\delta$. \cite{Nakanishi93}

The spin-freezing temperatures $T_{f}$ of the alloys studied have been
obtained from  ac-susceptibility  measurements \cite{Obrado98b}.    As
shown in Table  \ref{SAMPLES}, in all  the samples studied $T_{f}$  is
below  the martensitic transition.  We  also present the values of the
paramagnetic Curie temperature $\theta_c$  and the temperature $T_{M}$
of      the   structural    transition    taken   from    the     same
Ref. \onlinecite{Obrado98b}.  Concerning  $\theta_c$, note that it  is
positive for the three samples, which indicates the mean ferromagnetic
character of the interactions.

\begin{figure}
\epsfig{file=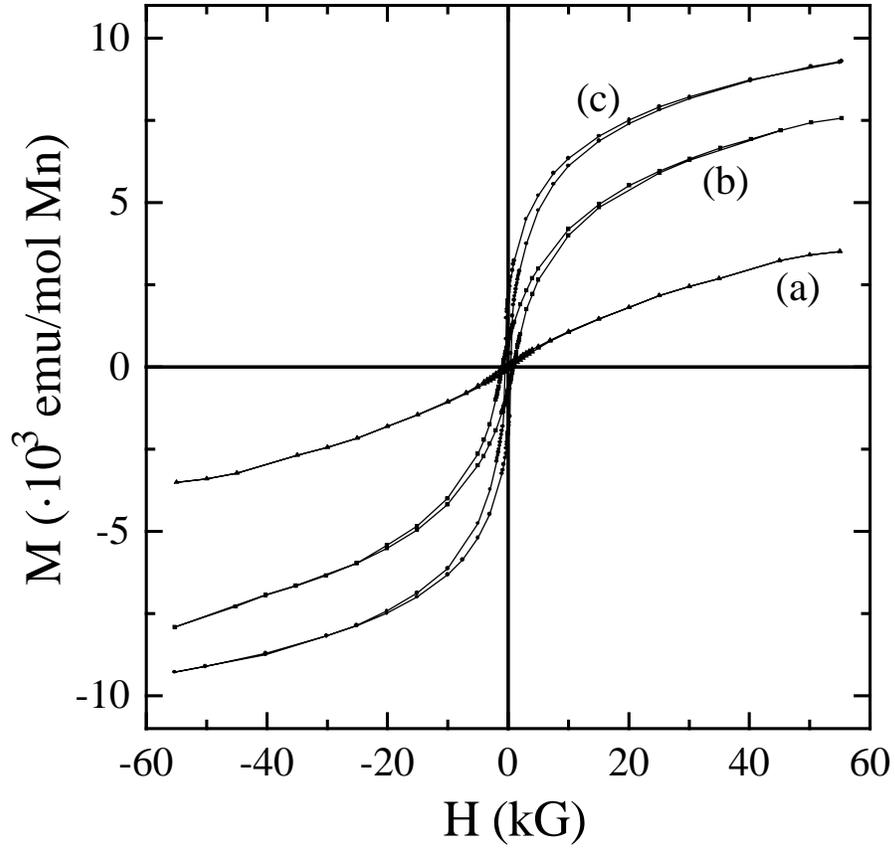,width=12cm}
\caption{Hysteresis   cycles  at  $T=5$  K   at   the  three different
compositions studied: $x=0.12$ (a), $x=0.25$ (b) and $x=0.36$ (c).}
\label{FIG2} \end{figure}

Magnetization measurements have been carried  out, at low field  rates
($<10$ Oe/s) using a SQUID magnetometer  enabling applied fields up to
5 T (the relative error  in the field  control is $\pm  0.1$ \%).  The
hysteresis observed in the  glassy phase reflects  the non-equilibrium
state  of the system characterized  by a logarithmically slow response
of  the magnetization to an  applied field.  Limiting hysteresis loops
have  been measured  after  a zero-field cooling  (ZFC) and subsequent
(isothermal)  field  sweep  between $-5$  T and  $+5$  T at  different
temperatures (controlled within $\pm 0.0 1$ K). In figure \ref{FIG2} we
show the hysteresis  cycles obtained  at $5$  K in  the three  samples
studied.  Loops  are   symmetrical and display small   coercivity  and
remanence.  It is  worth mentioning that  for the sample with $x=0.25$
($6.3\%$ Mn), the hysteresis loop has also been measured after a field
cooling (FC).  The  cycle obtained is  identical from the one measured
after a  ZFC.   This proves  the absence  of global  uniaxial exchange
anisotropy in the polycrystalline samples considered.

From   the results  given  in Fig.\  \ref{FIG2} it  is  clear that the
magnetization can be supposed to be  composed of an irreversible and a
reversible contributions:\cite{Prejean80}

\begin{equation}
M=M_{rev}+M_{irr}
\label{Magnetization}
\end{equation}

Only  the irreversible term contributes  to hysteresis.  Actually, the
reversible  contribution  must be ascribed to   the competition of the
external field with minoritary  antiferromagnetic bonds in the system,
and can be assumed to be linear with the magnetic field:

\begin{equation}
M_{rev}=\chi (T)H
\label{Mrev}
\end{equation}

From  linear    fitting, we have  determined   $\chi$  for  the alloys
investigated.   The  values obtained are:   $\chi  = 0.029  \pm 0.010$
emu/(mol\,Mn Oe), $\chi = 0.041  \pm 0.010$ emu/(mol\,Mn Oe) and $\chi
= 0.023 \pm 0.010$ emu/(mol\,Mn Oe)  for the samples with $x=0.12$ ($3
\%$  Mn),   $x=0.25$    ($6.3  \%$  Mn)  and   $x=0.36$    ($9\%$  Mn)
respectively.  Therefore,   within  errors,   $\chi$  can  be  assumed
independent of $x$,   at least in  the range  of compositions studied.
After  subtraction  of the    reversible  term in    the magnetization
measured,  we have estimated  a saturation magnetization $M_s$.  It is
represented (per Mn atom) as a function of $x$ in Fig.\ \ref{FIG3}. We
obtain that  $M_s$  increases linearly  with $x$.  The  value of $M_s$
corresponding to   the stoichiometric  Cu$_{2}$AlMn   alloy ($M_s(x=1)
\equiv \mu  \simeq  4.1 \mu_B$) has  been  taken from  the  literature
\cite{Webster88}.

As  temperature  is increased   the  width of the   cycle (coercivity)
decreases and, as expected,    we have not observed  hysteresis  above
$T_{g}$.  In  figure \ref{FIG4} we  show the effect of  temperature on
hysteresis by plotting the  central  part of  the hysteresis loops  at
three different temperatures.  In all cases the  cycles are smooth and
no  discontinuities (or avalanches)  have  been detected.  However the
cycle corresponding to $x=0.36$ obtained at $T=5$ K, shows practically
a complete reversal  of the remanent magnetization  in a very  reduced
interval of fields.   This changes  the shape  of the hysteresis  loop
which becomes sharper in its central part.

In order  to quantitatively characterize  the shape of  the hysteresis
loops, we have represented in  Fig.\ \ref{FIG5} the remanence $M_{r}$,
the coercivity $H_c$ and the area of  the loops $E$,  as a function of
temperature for the three different compositions studied.  The area of
the loops, which gives the dissipated energy per cycle, decreases with
$T$ and increases with  the Mn content  $x$.  Notice also that for the
sample with $x=0.36$ the coercive field does not  change for $T<10$ K.
This fact is associated with the change in the shape of the hysteresis
loop outlined in the previous paragraph.

We  argue  that the observed  change  in  the  hysteresis loop can  be
related to the fact that, as has recently been shown \cite{Obrado98b},
a spin-glass phase transition occurs for samples with $x<0.3$, but not
for samples with higher Mn concentration.  For these compositions, the
cusp in the magnetic susceptibility curve as a function of temperature
is related to  the  freezing  of  ferromagnetic clusters, being    the
high-temperature  phase superparamgnetic and the low-temperature phase
what we   will   call  in this   article   a  ferromagnetic spin-glass
\cite{aclarir}. Therefore we suggest that the change in the hysteresis
loop may be related to a phase transition which  has the origin on the
different ground  state properties of the system  as $x$ is increased.
Actually the value of $x$ at the  transition point, should be close to
the percolation limit for the  system  studied.  It would be  expected
that remanent  magnetization reverses at a  single value of  the field
(large avalanche).  The fact that this does not strictly occur is due,
as we    will   discuss in   the  next  section,  to   the
polycrystalline character of the samples studied.

We have  also studied first-order  reversal cycles within the limiting
loop obtained as follows: starting from $H=-5$ T the external field is
increased  to    a  value $H_{max}$    ($H_{c}<H_{max}<5$  T),  before
saturation.  At this point   the field evolution  is reversed  back to
$H=-5$  T.  For    different $H_{max}$   one  obtains  a   family   of
trajectories, some of which are displayed in Fig.  \ref{FIG51} for the
sample with $x=0.25$ and $T=5$ K. It is worth noting that in all cases
the decreasing   reversal trajectories join tangentially  the limiting
cycle.     The   shape of these   reversal     trajectories depends on
$H_{max}$. In the inset of Fig. \ref{FIG51} we show the coercive field
of the reversal trajectories $H_c^{part}$ as  a function of $H_{max}$.
Note that for the sample with $x=0.36$ (circles), $H_c^{part}$ remains
approximately constant   with  $H_{max}$ when   aproaching $H_c$  from
above.   This  is indicative  of   the fact   that  the  reversal   of
magnetization occurs at a fixed field for high $x$.

An interesting feature observed in a number of systems exhibiting FLFO
phase transitions is the return point memory effect.  This property is
well illustrated by analysing the characteristics of complete internal
hysteresis loops.  These internal loops are obtained when, coming from
$H=-5$ T and being in a state corresponding to a  field $H_{1}<5$ T on
the  limiting cycle, the   field  evolution  is reversed  back   until
reaching $H_{2}<  H_{1}$, and later  is increased up again to $H_{1}$.
The  return point memory  effect is verified  if, after this path, the
system returns exactly to the same state that was left in the limiting
cycle, that is, if  it has memory of the  former state.  In this case,
the same memory effect extends  to any subcycle  within a given cycle,
and therefore this indicates that the state of the system can remember
an entire hierarchy of  turning points  in  its past evolution.   This
phenomenon has been  observed in  some, but by  no  means all, systems
with FLFO phase transitions.

We have performed such complete internal cycles by reversing the field
at different states in the limiting cycle.   Typical results are shown
in Fig.\ \ref{FIG6}.  It is observed that after performing an internal
loop  and returning   to  the  former  field,  the  magnetization  has
increased.  By performing internal  cycles at different sweeping field
rates (in the positive  and negative applied field  regions) , we have
checked that this  increase  of  the magnetization  is   significantly
larger than  the  increase  expected   from logarithmic   relaxational
effects.  This  enables to  conclude  that, in our system,  the return
point memory property is not satisfied.

\section{MODELING}

In order to gain understanding  of the observed experimental phenomena
we develop, in this section, a model containing two basic ingredients:
randomness of the lattice position of the magnetic atoms and competing
interactions.  These two ingredients are crucial  for the existence of
spin-glass phenomena \cite{Fischer91}.     The proposed   model is   a
modified version of the diluted Ising model at $T=0$.\cite{Tadic96}

We only consider sublattices  $\alpha$  and $\beta$ as possible  sites
for Mn   atoms (see Fig.  \ref{FIG1}).  Therefore   we define a simple
cubic lattice  \cite{martensita}  with $i=1,..,N$  ($N=L^3$) sites and
with  periodic boundary conditions and  lattice spacing  $a$.  On each
site we define a  variable $c_i=1,0$ which  indicates the presence (1)
or absence (0) of Mn on that site, and an  Ising variable $S_i= \pm 1$
which  accounts for two  possible magnetic  states of  Mn atoms.   The
hamiltonian of the model is written as:

\begin{equation}
{\cal H} = \sum_{ij} J_{ij} c_i  c_j S_i S_j - \mu  H \sum_{i} c_i S_i
\; ,
\end{equation}

\noindent where the  first term is a  sum over site pairs, $J_{ij}$ is
the exchange  energy, $H$ the  external field and  $\mu$ the effective
magnetic  moment per atom. A  number $N_{Mn}=\sum_{i} c_i$ of Mn atoms
are randomly placed as follows: a fraction $(1-p) N_{Mn}$ is placed in
sublattice  $\alpha$ while  the rest $p  N_{Mn}$   is ``misplaced'' in
sublattice $\beta$.  Values of $p$ ranging from $0$ to $0.5$ have been
considered.  The concentration of Mn is  given by $x=N_{Mn}/(2N)$. The
exchange   constants  $J_{ij}$   are distance   dependent.    For  the
stoichiometric alloy, $J_{ij}$ have been fitted from inelastic neutron
scattering experiments \cite{Tajima77}.   The  fitted values  are  the
ones  corresponding to  neighbors  of 2nd,  4th, 6th and  7th order in
Table \ref{JIJ}.  We assume that  $J_{ij}$ do not change significantly
when  altering the Mn     content.  Nevertheless  we  still have    to
interpolate the values  of  the  $J_{ij}$ corresponding  to  distances
which are not present in perfectly ordered  stoichiometric alloys.  To
do that we have fitted a standard  RKKY dependence to the experimental
$J_{ij}$:

\begin{equation}
J(r)= A \frac{\cos(B (r/a))}{(r/a)^3}
\end{equation}

\noindent  where $B$ is proportional to  the Fermi moment $B=2 k_F a$.
Since  it is difficult  to fit  such  an oscillating function we  have
first checked different  values  of $B$ in  order  to get the  correct
signs of  the  $J_{ij}$  found  experimentally.  The signs   are  only
correctly    fitted  for values   of      $B$ around   $\sim   19.64$.
\cite{Aclariment}  A  standard   least square fitting,  then,  renders
$A=-2.58$  meV.   Using this  fit we  then   interpolate the values of
$J_{ij}$ for  the neighbors of order  $1$st, $3$rd and $5$th indicated
in Table \ref{JIJ} with asterisks.

A strong antiferromagnetic  interaction between n.n.  sites  (that can
only occur due to misplaced Mn  atoms) is obtained. This is consistent
with the  antiferromagnetic  character found  in  $\gamma$-Mn with fcc
structure \cite{Gibbs85} (lattice  parameter $<2a$).  Nevertheless  we
have checked that the mean character of the interactions (which can be
estimated as a sum $\sum n_{ij} J_{ij}$ over all Mn pairs $n_{ij}$) is
always ferromagnetic for all values of $x$ and $p$. This is consistent
with   the   positive paramagnetic     Curie   temperatures   obtained
experimentally (see Table \ref{SAMPLES}).

For  the simulations we consider the  exchange energy for next-nearest
neighbors  $J_2=0.744$ meV  as the  energy  unit.  We define a reduced
hamiltonian as:

\begin{equation}
{\cal  H}^* = \sum_{ij}  J_{ij}^* c_i c_j  S_i  S_j + H^* \sum_{i} c_i
S_i\;,
\end{equation}

\noindent where $J_{ij}^* = J_{ij}/  J_2$ and $H^* =  \mu H / J_2$. We
define the reduced magnetization as:

\begin{equation}
m^* = \frac{\sum_i S_i}{N/2}
\end{equation}

With this  definition,   the  maximum reduced   magnetization for  the
stoichiometric alloy is $m^*=1$.

Hysteresis cycles ($m^*(H^*)$) are  obtained by starting with  all the
spin variables  $S_i=-1$ and the  external field  $H^*=- \infty$.  The
field is then slowly increased  until a spin becomes locally unstable,
i.e.  there  is  a decrease of  energy  if  it is  flipped.   Then the
external field   is  stopped, the  spin is  flipped  and an  avalanche
starts: more spins   may become  unstable.  These  are  simulatenously
flipped, which  may trigger some  new unstable spins,  and so on until
the  avalanche stops.  The  number  of steps necessary  to reach  this
stable situation defines  the avalanche duration and the corresponding
magnetization change the avalanche size.   Then the external field  is
increased     again.      This    is    the    so-called   synchronous
dynamics. \cite{degeneration}

\subsection{Results}

We have studied lattices with $L=32$ and for  many runs, averages over
$50$ different initial random  atomic configurations have  been taken.
We have systematically studied systems  with concentration $x= 0.125$,
$0.25$, $0.375$, and $0.5$ and disorder values $p=0.1, 0.2, 0.3, 0.4$.
Examples of the hysteresis cycles are shown in  Fig.\ \ref{FIG7}.  For
a fixed value of  $p$, the hysteresis loop  becomes sharper as $x$  is
increased, while  for a fixed value  of $x$ the loop  becomes smoother
when increasing $p$.  Note  also that for  high values of $p$ and  $x$
the  loops  become antiferromagnetic-like with long  tails  and with a
tendency to split  into two subloops.   In all cases, when  looking in
detail  at  the loops, they consist  of  a sequence  of discrete jumps
(avalanches)   that join metastable  states.   This  is illustrated in
Fig.\  \ref{FIG8}.    Due   to  the  existence  of   antiferromagnetic
interactions,  inverse  avalanches, decreasing  the magnetization when
increasing  the field, may occur.  Concerning   this point, one should
remember that the evolution is controlled by  a local (and not global)
equilibrium condition.

In order to study  in more detail  the evolution from smooth to sharp
hysteresis  cycles we have  studied   the statistical distribution  of
avalanche sizes.  Sharp loops will exhibit  few large avalanches while
smooth  loops will   exhibit  small  avalanches  only.  The  avalanche
distributions   have  been computed  by  averaging   over 50 different
realizations  of the random distribution  of Mn  atoms.  The resulting
histograms are presented in Fig.\ \ref{FIG9} for different values of p
and x.  We find that, in  this p-x diagram,  a critical line exists at
which avalanches  of all sizes are  observed and  the distributions are
power-law.  This behavior  has been analyzed following  the techniques
developed for the study of avalanches in  a previous work on the study
of   FLFO    phase transitions  in   the    Random   Bond Ising  Model
\cite{Vives94b}.

The critical line can  be found by  studying  two quantities: (i)  the
normalized average    size  $\langle \Delta m_{max} \rangle$    of the
largest avalanche in a  cycle and (ii)  the average duration $ \langle
t_{max} \rangle$ of  the largest avalanche in  a cycle.  As an example
we show the  behavior of these two quantities  along a line with $p=0$
and   changing the concentration $x$ (Fig.    \ref{FIG10}) and along a
line with  $x=0.375$ and  changing  the disorder  parameter $p$  (Fig.
\ref{FIG11}).  The  phase transition  is  smoothed due to  finite size
effects,   which    can  be  corrected      as  explained   in   Ref.\
\onlinecite{Vives94b}.  Actually, estimations of the transition points
can be obtained  by locating, using  a polinomial fitting, the maximum
on  $\langle t_{max} \rangle $ and  the  inflection point on $ \langle
\Delta m_{max} \rangle $.  A systematic study of the behavior of these
two quantities as a function of $x$ and $p$ allows to obtain the phase
diagram  shown in Fig.  \ref{FIG12}.   The discrepancy between the two
estimations  of   the   critical   line    can   be  attributed     to
finite-size-effects. We find critical behaviour, depending on $p$, for
$x \geq 0.3$.

\begin{figure}
\epsfig{file=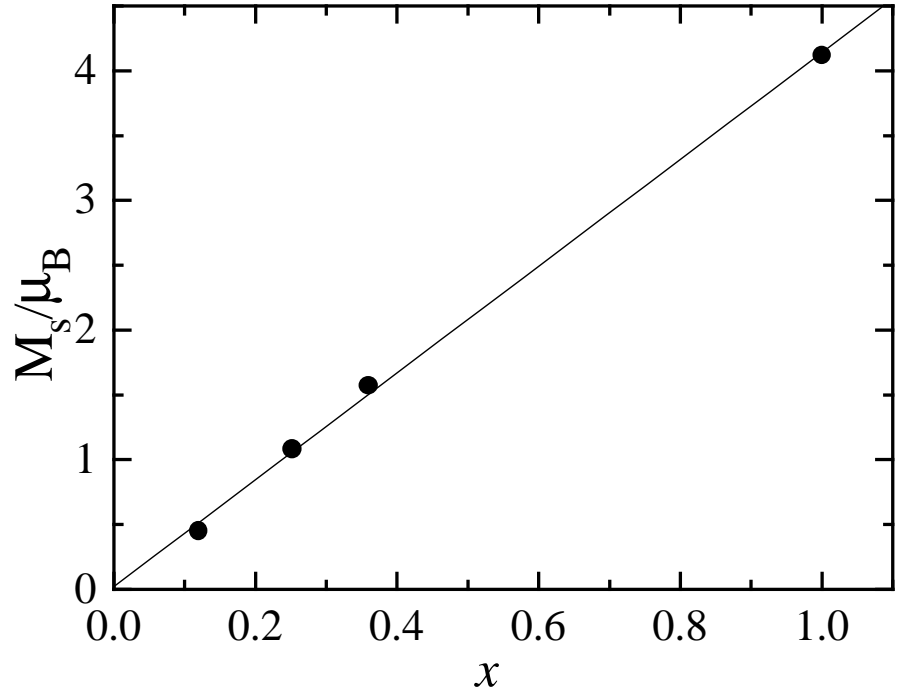,width=12cm}
\caption{Saturation magnetization as a function of $x$. The continuous
line shows a linear fit.} \label{FIG3}
\end{figure}

\begin{figure}
\epsfig{file=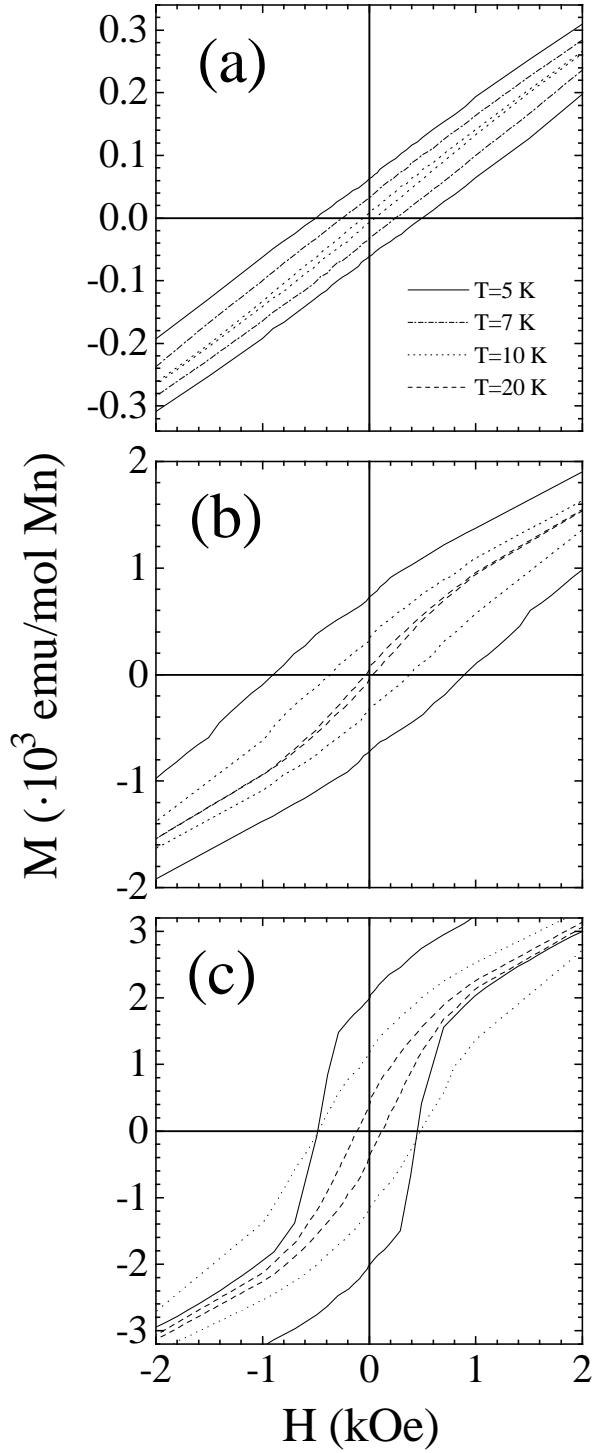,width=8cm}
\caption{Central  part of   the hysteresis cycles   for three diferent
temperatures and compositions: $x=0.12$ (a), $x=0.25$ (b) and $x=0.36$
(c)}
\label{FIG4}
\end{figure}

\begin{figure}
\epsfig{file=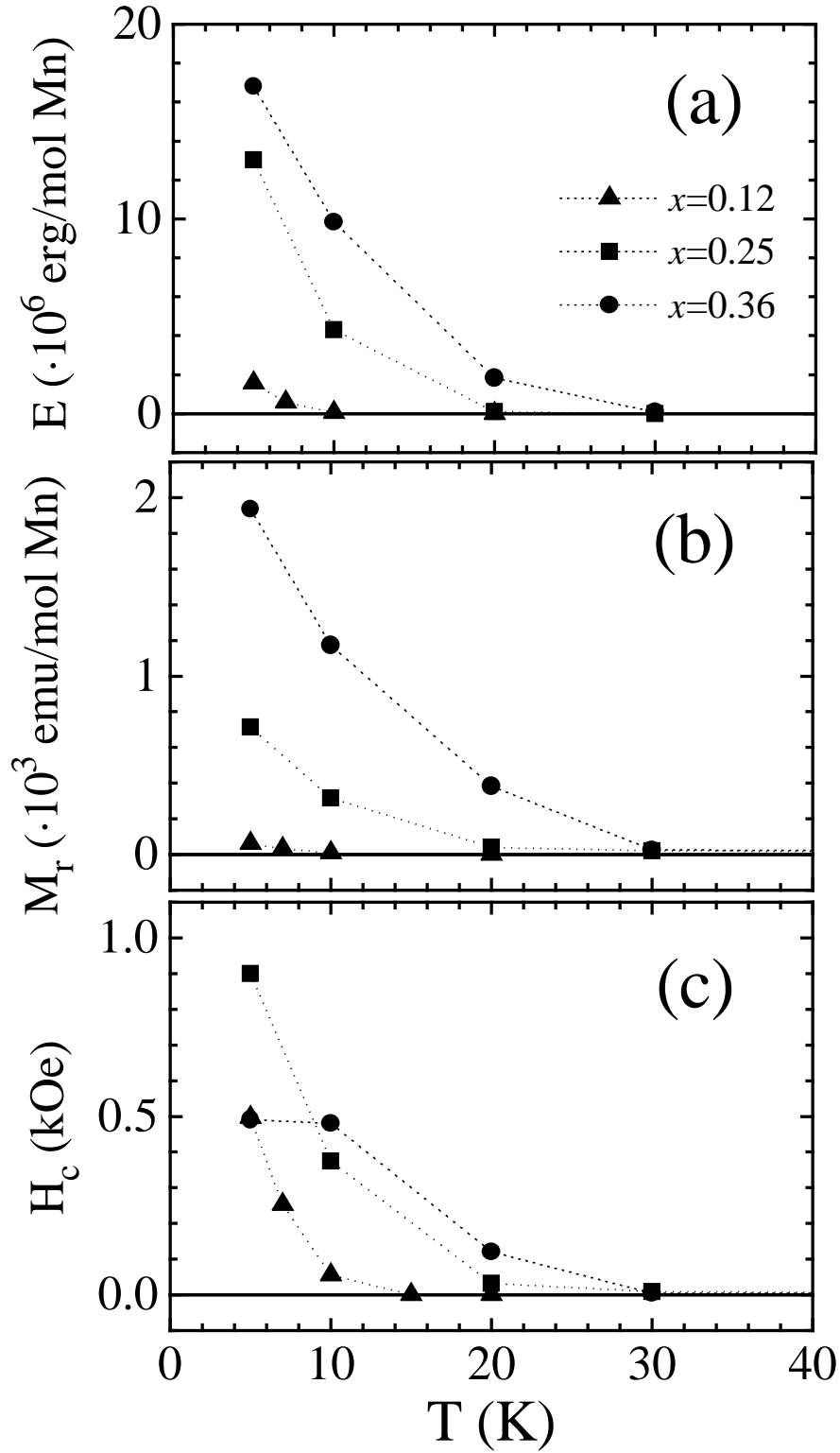,width=12cm}
\caption{Hysteresis  cycle  properties  versus  temperature  for three
different  compositions: dissipated energy (a), remanent magnetization
(b) and coercive field (c). } \label{FIG5} \end{figure}

\begin{figure}
\epsfig{file=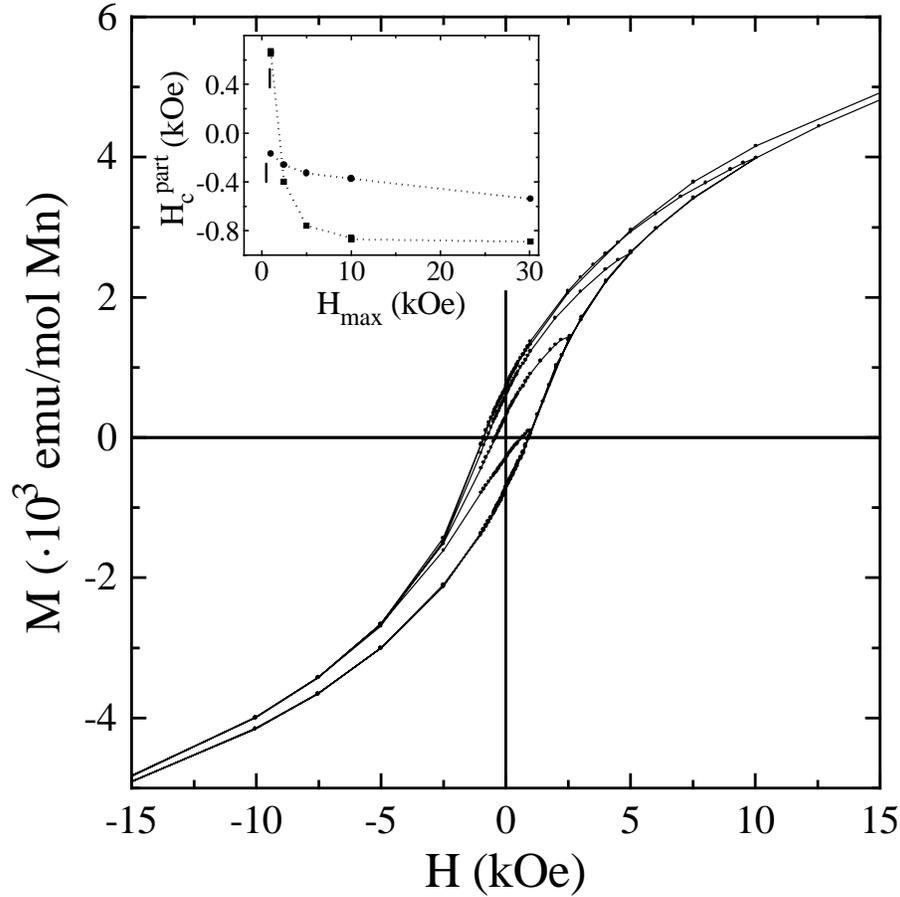,width=12cm}
\caption{Partial   loops corresponding to  a  sample with $x=0.25$ and
$T=5$  K.  The    inset shows  the   behavior  of  the coercive  field
$H_{c}^{part}$   as a  function  of $H_{max}$   for   the samples with
$x=0.25$ (squares) and $x=0.36$ (circles).  The vertical bars show the
values of the coercive field for the limiting loops in each case.}
\label{FIG51} 
\end{figure}

It is difficult  to locate on this $p-x$  diagram the position  where this
phase transition will be expected to occur in the Cu-Al-Mn system,
 since the value of
$p$ for our samples is unknown.  Actually we can  only have some rough
estimations of  the $p(x)$ dependence in  the samples studied.  First,
it   is possible to obtain  $p$  from Monte Carlo   simulations of the
atomic   ordering process in a   realistic lattice  model for Cu-Al-Mn
presented  in Ref.  \onlinecite{Obrado98}.   The  model  considered in
that   paper contains only   interactions to first  and second nearest
neighbors and its parameters have been  fitted from the order-disorder
transition temperatures in the real experimental system.  The behavior
of $p$ as a function of $x$ obtained from that model is represented by
the two dotted lines in Fig.  \ref{FIG12}.  The lower line corresponds
to the values of $p$ at $T=0$, and the  upper represents the values of
$p$  at  temperatures   just below    the order-disorder   transition.
According to this result, the actual values  of $p$ would be somewhere
in the middle of  these two dotted  lines.  The intersection  of these
lines with the phase transition (open symbols) would give the critical
concentration where  the change in the  hysteresis  cycle is expected.
However,  these   estimations  of  $p(x)$   correspond to  equilibrium
configurations, but kinetic effects  (samples have been quenched  from
high $T$)    could  modify  the   amount  of  disorder  substantially.
Moreover,  the model used for the  MC simulations does not contain any
magnetic interaction term, which could  have some relevance in the  Mn
ordering process.  A second estimation can be obtained from the linear
behavior of the saturation magnetization per atom $M_s$ with $x$ shown
in  Fig.\ \ref{FIG3}.  Note  that the system  contains two kinds of Mn
atoms: the minoritary  misplaced  ones (in  sublattice $\beta$) that
exhibit antiferromagnetic  interaction with its nearest neighbours and
the  majoritary  ones   (in  sublattice $\alpha$)  with  ferromagnetic
interaction with its  second neighbours.  Therefore, when reaching the
remanence  point by decreasing the applied  field from saturation, the
misplaced Mn   atoms will present  inverted magnetic  moment.  Thus, a
naive approximation  is $M_s = \mu \left  ( (1-B)(2p -1)^2 +  B \right
)$, where  $\mu$ is the magnetic moment  per atom and   $\mu B$ is the
saturation magnetization per atom   corresponding to a sample  with Mn
atoms completely disordered over the  $\alpha$ and $\beta$ sublattices
($p=0.5$).  Note that $M_s(p)$ is  symmetric arround $p=1/2$ as can be
expected.  The above equation combined  with the linear behaviour $M_s
\simeq \mu  x$  presented in Fig.   \ref{FIG3},  leads to a  parabolic
dependence   $   x  = \left    (  (1-B)(2p  -1)^2   +  B   \right ) $.
\cite{xpetita} Although the value of $B$ is unknown, it should be very
small.  At least smaller than the value $M_s = 0.1 \mu $ found for our
sample with  $x = 0.12$.  We can,  therefore, sketch this behaviour on
the $p-x$ diagram shown in Fig.  \ref{FIG12}.  The intersection of the
two parabolas (corresponding to $B=0.1$ and $B=0$) with the transition
line  found from   our  model gives  a   different  estimation of  the
transition point.

Taking  into account the above   arguments, from Fig.\ \ref{FIG12}  it
seems reasonable to assume that the transition is located at $x_c \sim
0.5 \pm 0.1$.  Although this value is  slightly above the experimental
estimation    $x_c \sim 0.3$, the  values   are  still close enough to
suggest  that,  given the similarity  of  the  effects observed in the
experimental and simulated hysteresis cycles, the phenomenon may be an
example of the kind of phase transition found in the model.

From the model developed in  the present work, we  can also study  the
internal hysteresis loops in order to  compare with the behavior found
experimentally.  Fig.\ \ref{FIG13} shows, as  an example, an  internal
loop for a  system    with  $x=0.25$ and $p=0.2$.    The   qualitative
agreement  with the experimental case  is extremely good,  and in both
cases an  increase  of the  magnetization is found  after the internal
loop.  This shows that the RPM property does  not hold.  Actually, two
conditions must  be satisfied for the  RPM property  to be fullfilled.
The first one  is the adiabatic character of  the system evolution and
the  second one  is the  so-called   no-passing rule (preservation  of
partial   ordering    of   metastable    states   by    the  dynamics)
\cite{Middleton92}.   The  former   condition  is  satisfied in    our
simulations    (synchronous dynamics)   and    approximately   in  the
experiments (slow  driving   rate).  Nevertheless, the   existence  of
antiferromagnetic  interactions producing inverse avalanches (see Fig.
\ref{FIG8}) destroys the second condition.

\subsection{Polycrystallinity}

The model  developed above describes the  behavior of a single crystal
with an  easy axis  of  magnetization  (EAM) that corresponds  to  the
direction   of the external field.     In a polycrystalline sample the
situation   is   quite      different:   grains present      different
crystallographic orientations with respect  to the applied field.  The
corresponding angular distribution of EAM results  on two main effects
on the hysteresis  cycles: (i) the cycles  become  smoother (hence the
avalanches are much more difficult  to detect) and (ii) the hysteresis
cycle changes its  shape.  We can estimate  this change in the case of
uniaxial anisotropy. Let $m_0^*(H^*)$ be the  cycle corresponding to a
single  crystal oriented with the EAM  along the external field $H^*$,
and let  $m^*(H^*)$  the   cycle  corresponding to  the   polycrystal.
Assuming that  there is a uniform angular  distribution of EAM  in the
polycrystal, we will have:

\begin{equation}
m^*(H^*) = \int \cos \theta \; m_0^* (H^* \cos \theta ) \; d \Omega\;,
\label{anis}
\end{equation} 

\noindent where  $ d \Omega $  is the differential  solid angle.  This
integral has to  be performed over all the  possible directions of the
EAM  in the sample.  This  is  not  straightforward  and requires  the
knowledge of the direction of the EAM.   For a Cu-Al-Mn single crystal
it is not  known.  Nevertheless we  can obtain an approximate solution
to Eq.\ \ref{anis} as:

\begin{equation}
m^*(H^*) \simeq  \int_0^{\theta_0}   \cos \theta  \;  m_0^*  (H^* \cos
\theta) \; 2 \pi \; \sin \theta d \theta \;,
\end{equation} 
  
\noindent  where the  factor  $2 \pi  \sin   \theta d  \theta$ is  the
differential of solid  angle   and $\theta_0$  is the  maximum   angle
between the field and the EAM.  For instance  we have $\cos \theta_0 =
\sqrt{3}/3$ for a   $(100)$  and $(111)$  EAM,  and  $\cos \theta_0  =
\sqrt{2}/2$ for  a $(110)$ EAM.  In the  extreme  case of a monoclinic
symmetry (a single EAM)  one will have  $\cos  \theta_0 =  0$.  Figure
\ref{FIG14} shows the results of this correction on a given hysteresis
cycle. As expected the jump at the  phase transition becomes broadened
and extends in a certain range of fields.

\section{DISCUSSION AND CONCLUSIONS}

In  Section II we  have presented low-temperature  measurements of the
hysteresis   cycles of Cu-Al-Mn   in   three samples of  different  Mn
content.   We have shown  that for temperature  $T \lesssim 5$K and Mn
content $x>0.3$ the loops exhibit a sharp reversal of magnetization in
their central  part.  We have proposed  that this change is related to
the existence  of a phase   transition from a  spin-glass  phase to  a
ferromagnetic spin-glass phase.

In order to  gain insight into  this phase transition we have proposed
(in section III) a model  for the study of  the magnetic properties of
this   system,    including the  basic    physical  ingredients of the
problem.  The model is   based on a    $T=0$ diluted Ising  model with
exchange interactions that have been fitted to the experimental values
corresponding to  stoichiometric Cu$_2$AlMn.  The model reproduces the
existence of magnetic disorder arising  from the non-stoichiometry and
the misplaced  Mn atoms,  as    found experimentally.  It   should  be
remarked that we have not  studied the exact equilibrium ground  state
of the model. Rather we have focused on  the hysteresis loops that are
obtained by  using  a synchronous dynamics  evolution  that drives the
system through a sequence of metastable states.

Although the shape  of the loops has not  been perfectly reproduced by
the model, some of  the relevant features of  the real loops have been
found on the simulated ones. This includes:

\begin{enumerate} 

\item   The change from   smooth to  sharp cycles  by  changing the Mn
content.   In  the simulations this   change  corresponds  to a  phase
transition, of the same kind  as the one  reported in the Random Field
\cite{Sethna93} and Random Bond  Ising models \cite{Vives94b},  and it
is  found at $x =  0.5  \pm 0.1$, which  is  close to the experimental
estimation $x \sim 0.3$

\item The existence  of long tails for  large fields. The experimental
cycles are not saturated even at fields of $5$  T.  From the model, we
have deduced that   these  tails  arise  (i)  from  the  existence  of
minoritary antiferromagnetic interactions and (ii) as a consequence of
the polycrystalline character of the  samples used in the experiments.
The magnetic  moments  subjected   to an  effective  antiferromagnetic
interaction are very diluted    and  do not exhibit   any  cooperative
behavior.   Consequently   they  show a   reversible   response to the
external   field.  From the  model, it  is  reasonable to suppose that
these  magnetic  moments are associated   to misplaced Mn atoms having
other Mn atoms as nearest neighbors.  Concerning the polycrystallinity
of  the samples, its main  effect is the  broadening of the hysteresis
due to  the averaging over  all possible crystallographic orientations
of the grains with respect to the external field.

\item  The fact that the  RPM effect is not  satisfied  has been found
both experimentally  and   in  the  simulations.  The  reason  is  the
existence of inverse    avalanches  associated with the   presence  of
antiferromagnetic  interactions.  In  these avalanches $dM/dH$ is
negative,  contrary    to the  stability     conditions  deduced  from
equilibrium thermodynamics. The possibility of such a result is due to
the fact that  the process considered is  not an equilibrium one (even
if it is very  close to a quasiestatic process).   In fact the  system
evolves through   a  series of  metastable  states separated  by large
energy barriers.  In  the  simulations metastability arises   from the
fact that  relaxation is  performed  according to the  local effective
field acting on each magnetic moment.

\end{enumerate}

\begin{figure}
\epsfig{file=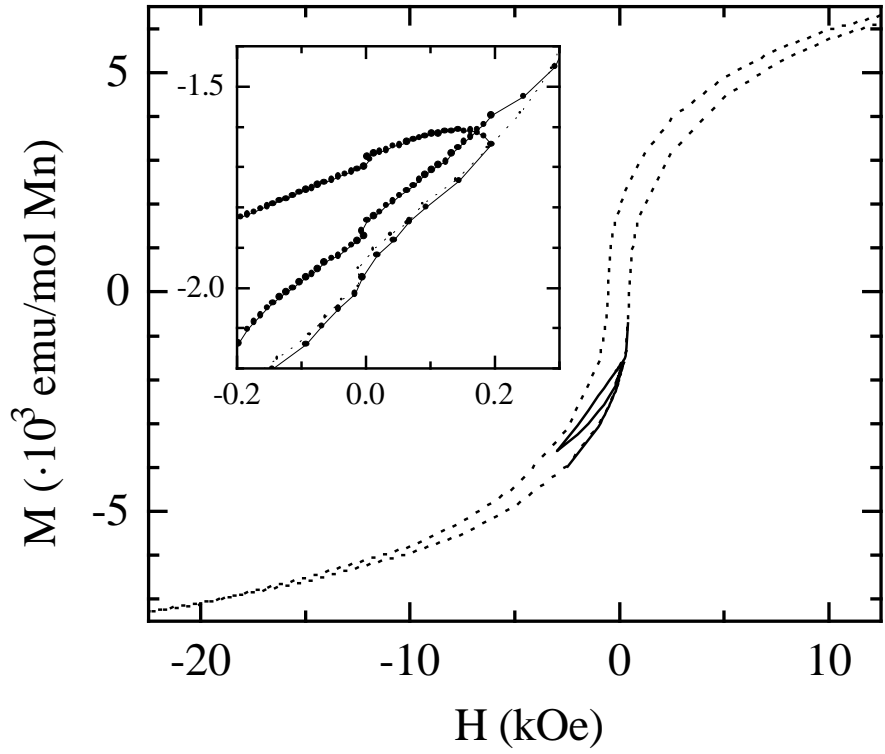,width=12cm}
\caption{Internal  loop corresponding to   a sample with $x=0.36$  and
$T=5$ K.  The inset shows a detail of the turning point.} \label{FIG6}
\end{figure}

\begin{figure}
\epsfig{file=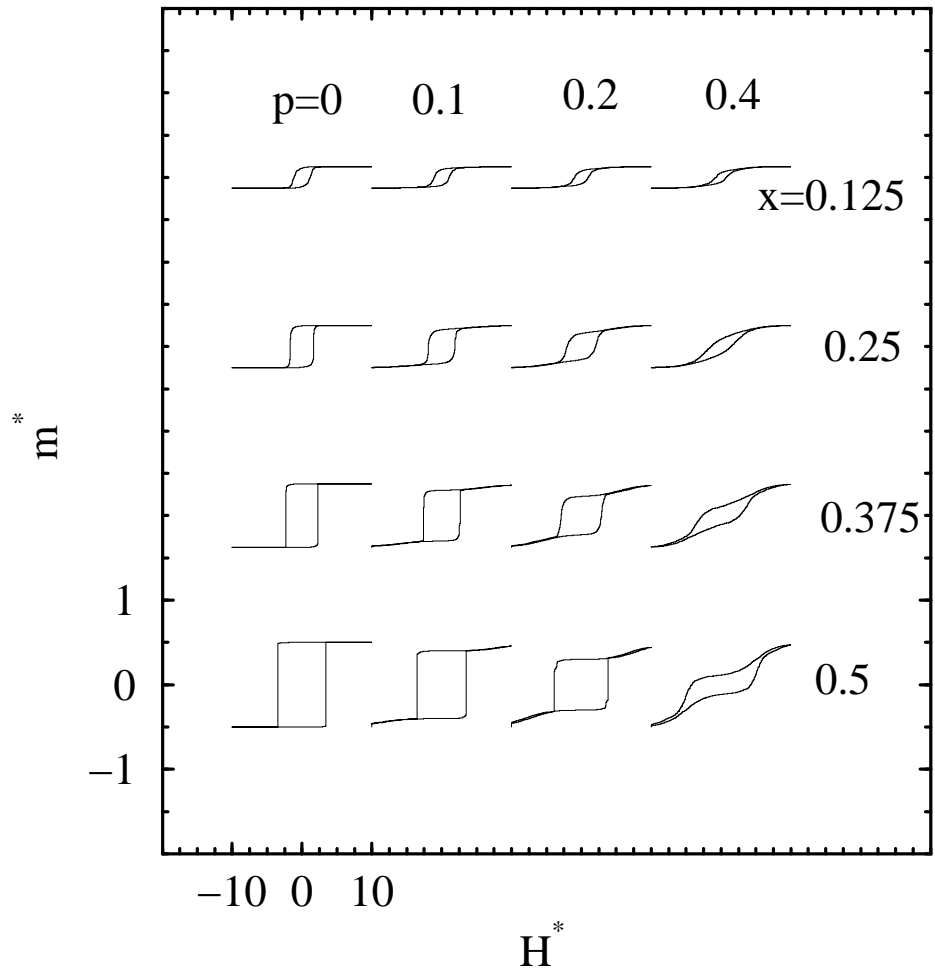,width=16cm}
\caption{Hysteresis cycles  for different  values  of $p$ and   $x$ as
indicated.} \label{FIG7}
\end{figure}

\begin{figure}
\epsfig{file=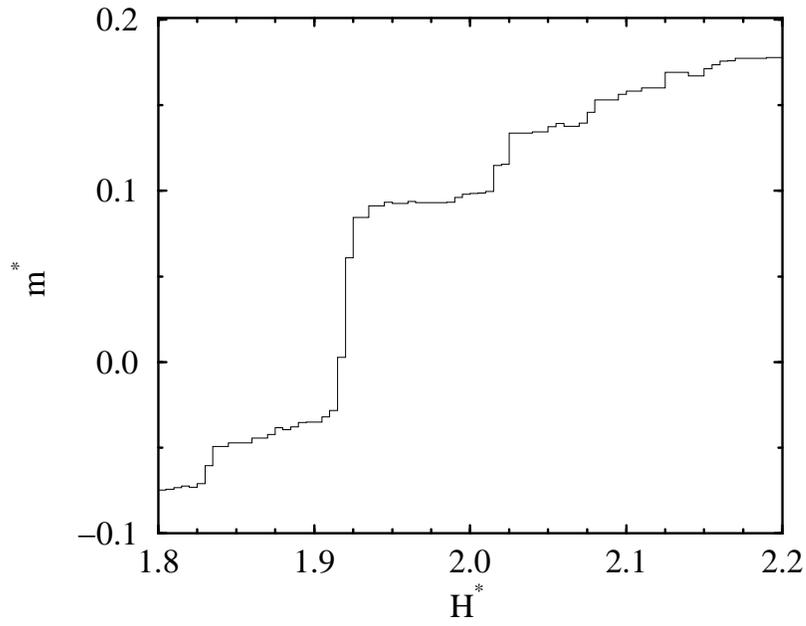,width=12cm}
\caption{Detail  of the avalanches for  a simulation with $x=0.25$ and
$p=0.10$.  Notice  the existence of  inverse avalanches,  e.g., around
$H^*=2.15$} \label{FIG8}
\end{figure}

An important question    to be answered   is why   avalanches are  not
observed in the experiments. It is clear that polycrystallinity limits
the maximum size of the avalanches to that of a  grain. Given that the
typical size  of the grains is  of the  order  of 100 $\mu$m,  and the
volume of the  sample  is $\sim$ 3  mm$^3$,  one will need a  relative
resolution much better than  $3  \times 10^{-4}$ in the  magnetization
measurement.  This  means a resolution  better  than 0.1 emu/(mol Mn),
while the actual resolution of our experimental system is of the order
of 10 emu/(mol Mn).  The inset  of Fig.\ \ref{FIG6} is illustrative of
this point.  It  would be very enlightening  to perform experiments in
single crystals, where we expect that avalanches could be observed.

Concerning the  phase transition found in  the model it is interesting
to discuss several points:

Firstly,   as mentioned before,   the phase  transition  found in  our
diluted  Ising  model (and   in   other similar diluted  Isind  models
\cite{Tadic96}) is of the  same  kind as  those  found for the  Random
Field \cite{Sethna93}  and  Random Bond Ising  models \cite{Vives94b}.
For these last two models it is known that the phase transition in the
hysteresis loop is associated  with an equilibrium ground  state phase
transition \cite{Ogielski86}.  It is reasonable to  think that for our
version of the site diluted Ising model,  there will also exist such a
ground state transition from   a  completely disordered ground   state
(small   $x$   region) to  a  ground      state with  a  non-vanishing
magnetization (large $x$ region).  Experimentally, there are evidences
that in   Cu-Al-Mn these two phases correspond   to a  true spin-glass
phase  and to a   ferromagnetic  spin-glass \cite{Obrado98b}.  In  the
model, whether the two phases  can be  catalogued as being  spin-glass-like 
would depend  on its thermal behavior.  This study is out of the
scope of the present work.

Secondly, the transition line in the $p-x$ diagram has been located from
the maximum of the  duration  and the inflection  of  the size of  the
largest  avalanche.    An additional estimation  can  be   obtained by
locating the points where the  distribution of avalanche sizes becomes
power  law.  Except  for  finite-size effects,  the  three estimations
should   be coincident.  Obtaining   the  transition points accurately
using the third  method is quite  difficult because it is necessary to
perform non-linear fits of a power-law with an exponential correction,
$N(s) \sim s^{-\tau} e^{  \lambda s}$, in   order to locate  the point
($\lambda= 0$) where the distribution  changes from super- ($\lambda >
0$)  to subcritical ($\lambda < 0$)  behaviour.   It is much easier to
evaluate the exponent $\tau$  characterizing the critical distribution
of  avalanches,   which  is  not affected  by   the small  exponential
corrections close to the critical line.  The obtained value is $\tau =
1.7 \pm 0.1$.  This value can be compared with the  values $\tau = 1.8
\pm 0.2$ and $2.0  \pm 0.2$ found  for the 3d-Random Field Ising model
and the 3d-Random Bond Ising model  \cite{Vives95}.  This agreement of
the  numerical values  reinforces   the hypothesis that  there  exists
universality in this kind of models exhibiting FLFO phase transitions.

\begin{figure}
\epsfig{file=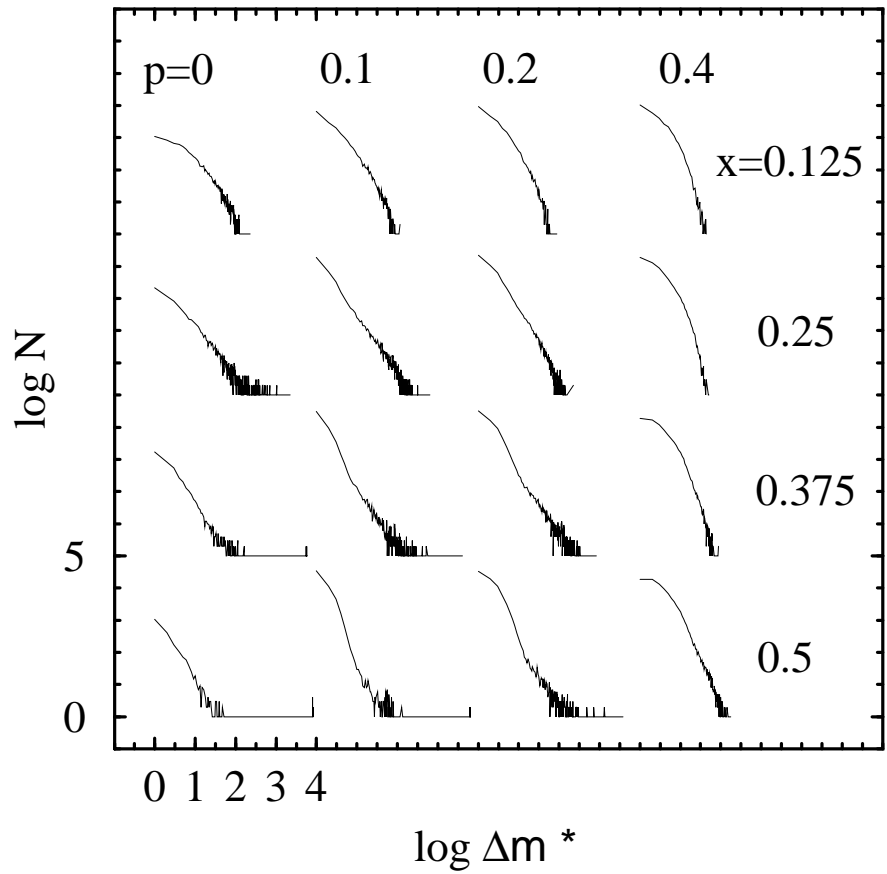,width=16cm}
\caption{Avalanche size distributions for  different values of $p$ and
$x$ as indicated.} \label{FIG9}
\end{figure}

\begin{figure}
\epsfig{file=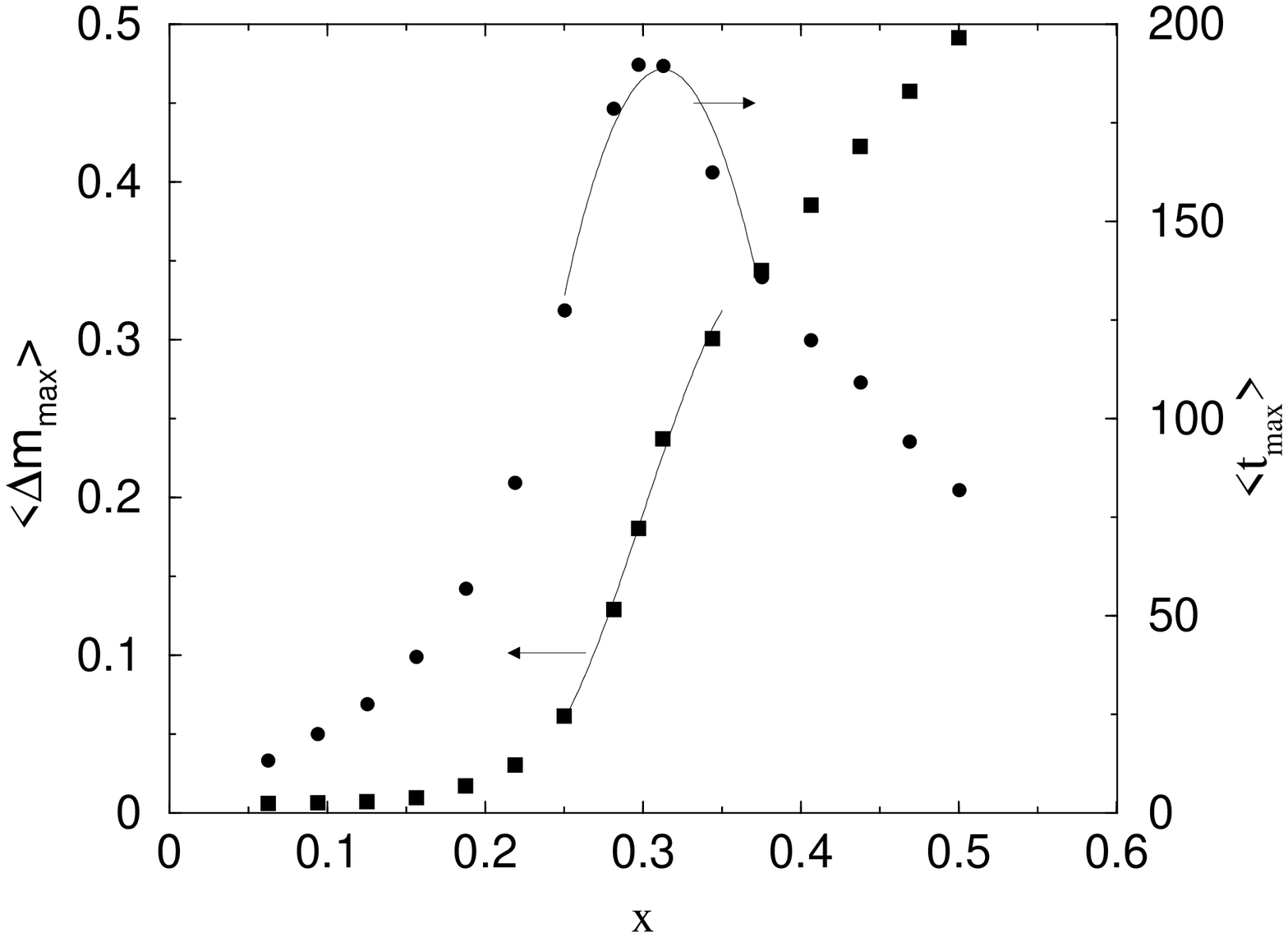,width=12cm}
\caption{Behavior of $\langle  \Delta m_{max}  \rangle$ and $  \langle
t_{max} \rangle$ as a function of $x$ for $p=0.0$.}
\label{FIG10}
\end{figure}

\begin{figure}
\epsfig{file=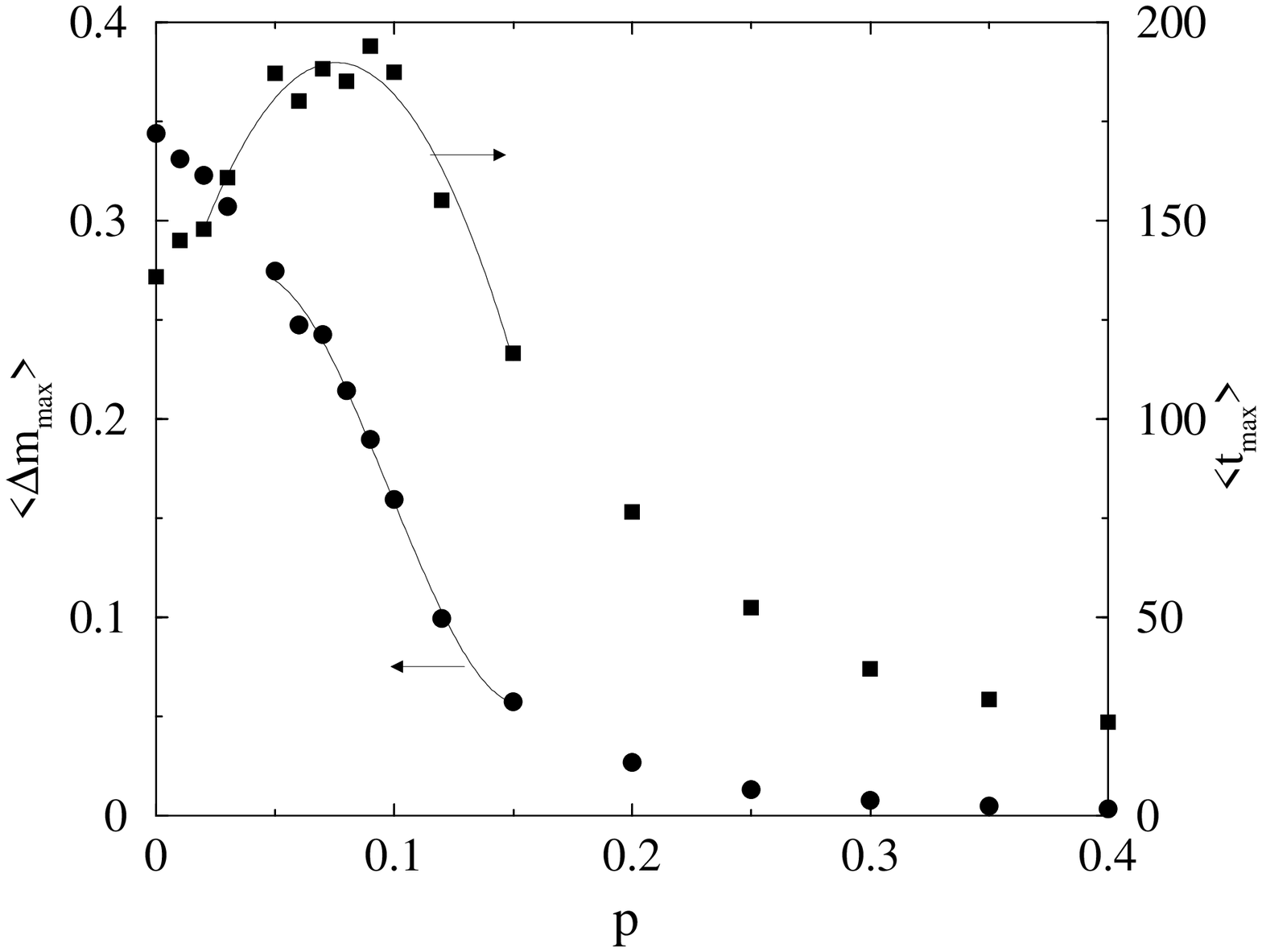,width=12cm}
\caption{Behavior of $\langle  \Delta  m_{max} \rangle$ and  $ \langle
t_{max} \rangle $ as a function of $p$ for $x=0.375$.}
\label{FIG11}
\end{figure}

\begin{figure}
\epsfig{file=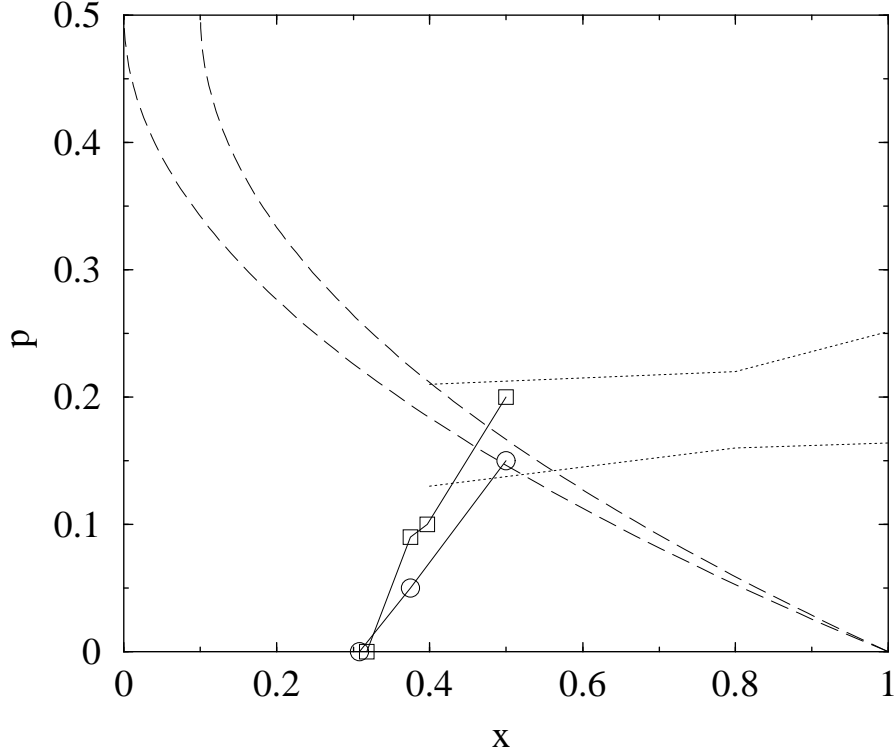,width=12cm}
\caption{$p-x$ diagram   showing   two different estimations   of  the
position of the FLFO phase transition.  Open squares correspond to the
positions of  the  maximum in $  \langle t_{max}  \rangle   $ and open
circles correspond to  the position   of  the inflection  points in  $
\langle \Delta  m_{max} \rangle  $.   Dotted lines  correspond  to two
estimations of  the $p(x)$ relation from  MC simulations, as explained
in the text. Long dashed lines correspond  to estimations based on the
behavior observed for the saturation magnetization.}
\label{FIG12}
\end{figure}

\begin{figure}
\epsfig{file=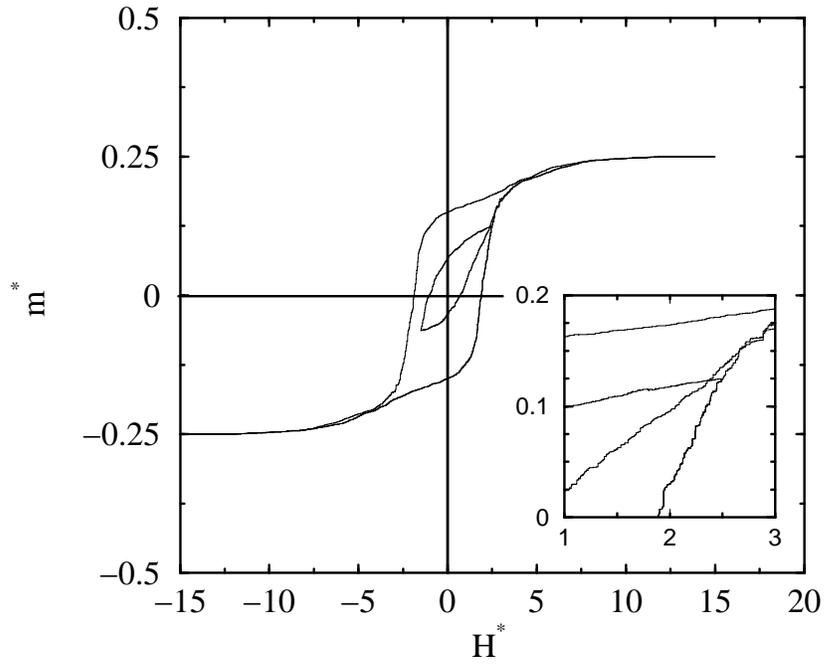,width=12cm}
\caption{Internal loops  showing the  absence of  RPM  property.  Data
correspond to a simulation with $x=0.25$ and $p=0.20$. The inset shows
the detail of the turning point.} \label{FIG13} \end{figure}

\begin{figure}
\epsfig{file=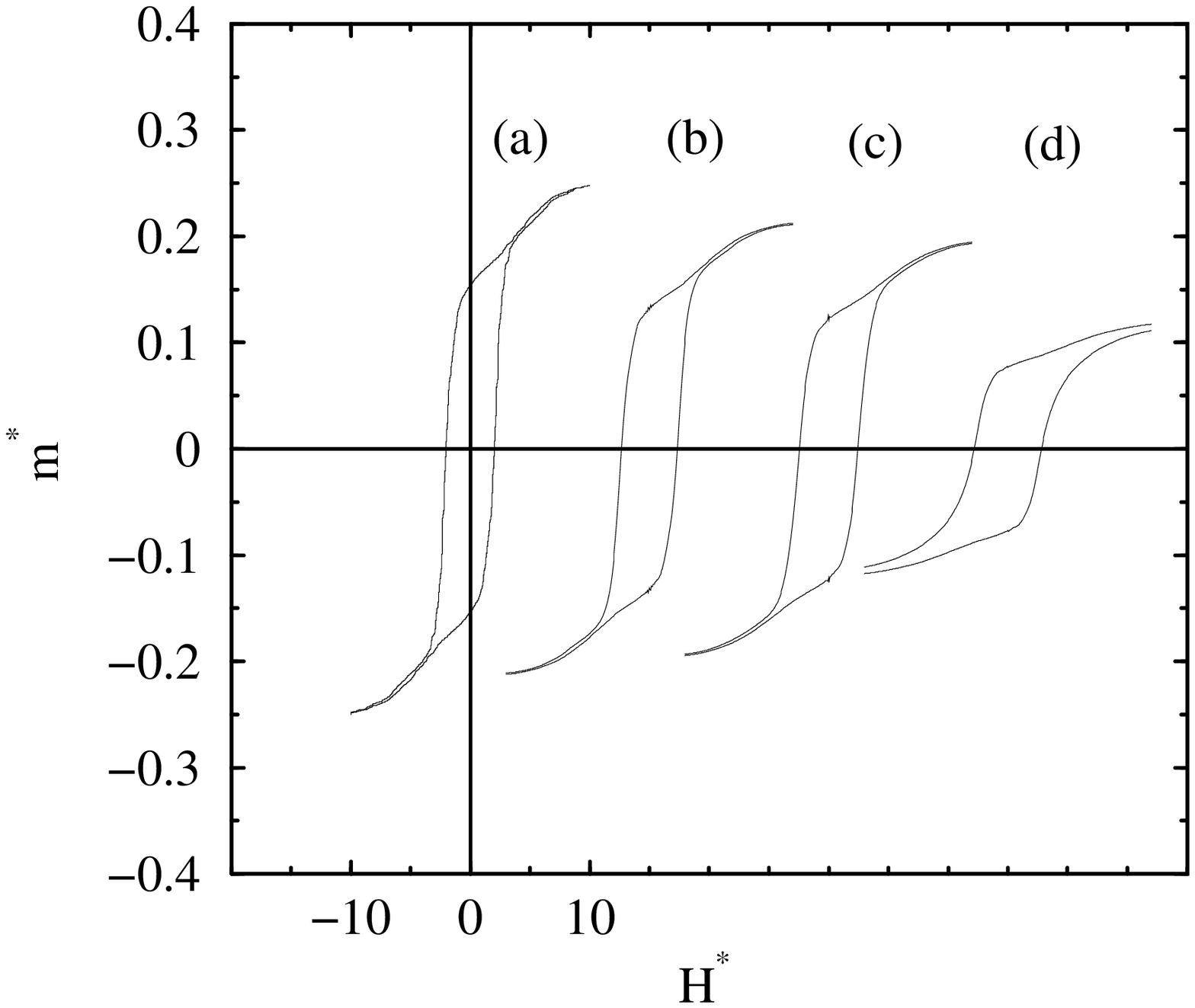,width=12cm}
\caption{Effect of the uniaxial anisotropy on  the hysteresis cycle of
polycrystalline systems. Data correspond  to the same simulation as in
Fig.  14.  The  different  cycles correspond  to: single crystal  (a),
$\protect \cos \theta_0   = \protect \sqrt{2}/2$ (b),   $\protect \cos
\theta_0 =  \protect \sqrt{3}/3$ (c)  and $\protect \cos \theta_0 = 0$
(d).}
\label{FIG14}
\end{figure}

Finally, a much  deeper question concerns the  nature of this critical
line.    Some  authors  have  stated  that  critical  distributions of
avalanches do not occur at a certain line but on a broad region due to
the existence of the so-called Self-Organized Criticality. This theory
has  been proposed for  externally  driven complex dissipative systems
with spatial and temporal degrees of freedom.  According to Bak et al.
\cite{Bak87} these systems   naturally   evolve to a   critical  state
characterized by avalanches with no  intrinsic time and length scales.
This results in power-law distributions for avalanches.  In this sense
the  results we have obtained  from  the model  point  in the opposite
direction:  we are  dealing  with a true  critical line.   Indeed, for
selected values of the  model parameters controlling the disorder  $p$
and $x$, the system exhibits the critical  behavior, as can be seen in
Fig.\ \ref{FIG9}.

\acknowledgements We acknowledge  Benjam\'{\i}n Mart\'{\i}nez for a helpful
colaboration and critical reading of the manuscript. We also thank Ll.
Ma\~{n}osa, J.  Ort\'{\i}n and  C.  Frontera for fruitful discussions.
This work has received finantial   support from CICyT (Spain)  project
number MAT98-0315, and CIRIT (Catalonia) project number SGR119.

\newpage

\begin{table}
\caption{Composition, characteristic spin-freezing temperatures, paramagnetic
Curie  temperatures and Martensitic transition temperatures of the samples 
investigated}
\label{SAMPLES}
\begin{tabular}{cccc}
Alloy composition  &  $T_f$ (K) &    $\theta_c$ (K) &   $T_M $  (K) \\
\tableline Cu$_{2.884}$ Al$_{0.996}$ Mn$_{0.120}$  & $15 \pm 1$ & $101
\pm 10$ & $390$ \\ Cu$_{2.804}$ Al$_{0.944}$ Mn$_{0.252}$ & $31 \pm 1$
& $138 \pm 10$ & $251$ \\ Cu$_{2.728}$ Al$_{0.912}$ Mn$_{0.360}$ & $45
\pm 1$ & $159 \pm 10$ & $133$
\end{tabular} \end{table}

\begin{table}
\caption{Exchange constants for the  different neighbors on the simple
cubic lattice formed by  sublattices $\alpha$ and $\beta$.  The values
with asterisk have been  interpolated as explained  in the text, while
the others have been  taken from Ref. {\protect \onlinecite{Tajima77}}
}
\label{JIJ}
\begin{tabular}{cccc}
neighbor & position & distance ($a$) &  $J_{ij}$ (meV) \\ \tableline 1
& $(100)$ & $1$ & $-2.098^*$ \\ 2 & $(110)$  & $\sqrt{2}$ & $0.744$ \\
3 & $(111)$ & $\sqrt{3}$ & $0.399^*$ \\ 4 & $(200)$ & $2$ & $0.296$ \\
5 & $(210)$ &  $\sqrt{5}$ & $-0.2343^*$ \\  6 & $(211)$ & $\sqrt{6}$ &
$0.355$ \\ 7 & $(220)$ & $\sqrt{8}$ & $-0.411$
\end{tabular}
\end{table}

\newpage

\newpage

\newpage

\newpage

\newpage

\newpage

\newpage

\newpage

\newpage

\newpage

\newpage

\newpage

\newpage

\newpage

\newpage

\begin{references}

\bibitem{Hysteresis}   G.   Bertotti,  {\it  Hysteresis in  Magnetism}
(Academic Press, N.Y.  1998).

\bibitem{Barkhausen19} A.  H.   Barkhausen,  Z.  Phys.  {\bf 20},  401
(1919).

\bibitem{Vergne81}          R.Vergne,       J.C.Cotillard          and
J.L.Porteseil. Rev. Phys.  Appl. {\bf 16}, 449 (1981).

\bibitem{Cote91} P.J.Cote and L.V.Meisel, Phys.  Rev.  Lett.  {\bf67},
1334 (1991);  J.S.Urbach, R.C.Madison   and J.T.Markert,   Phys.  Rev.
Lett. {\bf 75}, 276 (1995).

\bibitem{Vives94} E.    Vives,  J.  Ort\'{\i}n,  Ll.   Ma\~{n}osa,  I.
Rafols,  R.  P\'{e}rez -Magran\'{e} and A.   Planes, Phys. Rev. Lett.,
{\bf72}, 1694 (1994)

\bibitem{WuAdams} W. Wu and P.W.  Adams, Phys.  Rev.  Lett., {\bf 74},
610 (1995)

\bibitem{Lilly96}  M.P.    Lilly,  A.H.    Wotters,    R.B.   Hallock,
Phys. Rev. Lett., {\bf 77}, 4222 (1996)

\bibitem{Bertotti90} G.Bertotti and M.Pasquale, J.  Appl.  Phys.  {\bf
67}, 5255 (1990).

\bibitem{Vives94b} E.Vives and A. Planes. Phys. Rev.  B {\bf 50}, 3839
(1994).

\bibitem{Vives95} E.  Vives, J.    Goicoechea, J.  Ort\'{\i}n  and  A.
Planes, Phys. Rev E {\bf 52}, R5 (1995).

\bibitem{Sethna93}  J.P.Sethna,   K.Dahmen,  S.Kartha,  J.A.Krumhansl,
B.W.Roberts and J.D.Shore, Phys. Rev. Lett. {\bf 70}, 3347 (1993).

\bibitem{Babcock90}        K.L.Babcock       and       R.M.Westervelt,
Phys. Rev. Lett. {\bf 64}, 2168 (1990).


\bibitem{Carrillo98}  Ll.Carrillo, Ll.Ma\~nosa, J.Ort\'{\i}n, A.Planes
and E.Vives, Phys. Rev. Lett., {\bf 81}, 1889 (1998).

\bibitem{Kuskauer96} J.  Kushauer, R.  van Bentum, W.  Kleemann and D.
Bertrand, Phys. Rev. B {\bf 53}, 11647 (1996).

\bibitem{Cizeau97}  P.Cizeau, S.Zapperi,  G.Durin  and  H.G.  Stanley,
Phys. Rev.Lett. {\bf 79}, 4669 (1997).

\bibitem{Obrado98} E.  Obrad\'{o}, C.   Frontera, Ll.  Ma\~{n}osa, and
A. Planes, Phys. Rev. B, December 1998.

\bibitem{Webster88}   P.J.    Webster    and   K.R.A.   Ziebeck,    in
``Landolt-Bornstein New Series'', Vol.  III/19c, ed.   by O.  Madelung
(Springer-Verlag, Berlin 1988), p. 75


\bibitem{Tajima77}  K.    Tajima, Y.   Ishikawa,  P.J.   Webster, M.W.
Stringfellow, D.  Tochetti and K.R.A.   Ziebeck, J. Phys.  Soc. Japan,
{\bf 43}, 483 (1977)

\bibitem{Obrado98b} E. Obrad\'{o}, A.  Planes and B. Mart\'{\i}nez (to
be published).


\bibitem{Nakanishi93}  N.   Nakanishi,  T.   Shigematsu,  M.  Machida,
K. Ueda, K. Shimitzu and Y.  Nakata, ``Proceedings ICOMAT-92'', ed. by
C.M. Wayman and J. Perkins (Monterey, California 1993), p. 581

\bibitem{Prado98} M.O. Prado and F.C. Lovey, Acta mater., {\bf 46} 137
(1998)


\bibitem{Prejean80} J.    J.    Pr\'{e}jean, M.   J.    Joliderc   and
P. Monod, J. Phys. (Paris), {\bf 41}, 427 (1980).

\bibitem{aclarir}   By  ``ferromagnetic spin-glass''  we understand  a
phase which,  in absence of an applied  field, exhibits a large degree
of frozen disorder but with clusters with non-zero magnetization. Such
structures   have also been  called  cluster  glasses or mictomagnetic
phases.   See, for instance, C.M.Hurd,  Contemp.  Phys.  {\bf 23}, 469
(1982).


\bibitem{Fischer91}  K.H. Fischer and  J.A.  Hertz, {\it Spin Glasses}
(Cambrige University Press, Cambridge 1991).

\bibitem{Tadic96} B.Tadi\'c, Phys. Rev. Lett. {\bf 77}, 3843 (1996).



\bibitem{martensita} Actually   the structure  of  the  system  in the
martensitic phase  is    monoclinic  but, due to   the   diffusionless
character of  the  martensitic transition, the atomic   environment is
very similar in both the bcc and martensitic phase.

\bibitem{Aclariment} Such a value can be obtained by assuming a number
of conduction electrons per Mn atom $n_c=8$.

\bibitem{Gibbs85} P.Gibbs, T.M.Harders  and J.H.Smith, J. Phys. F {\bf
15}, 213 (1985).

\bibitem{degeneration} Since there are antiferromagnetic interactions,
inverse  avalanches may  occur  and,  in few  occasions, a  degenerate
situation may  appear with two energetically equivalent configurations
which successively alternate. In such cases we decide randomly which is
the stable one and proceed with the field increment.


\bibitem{xpetita} It    is   clear from   this   expression  that  the
approximation is not valid for very low values of $x$.

\bibitem{Middleton92} A.A.Middleton, Phys.  Rev.   Lett. {\bf 68}, 670
(1992).


\bibitem{Ogielski86} A.T.Ogielski, Phys. Rev. Lett. {\bf 57}, 1251 (1986);
M.R.Swift, A.J.Bray, A.Maritan, M.Cieplak, and J.R. Banavar, Europhys. Lett.
 {\bf 38}, 273 (1997).

\bibitem{Bak87} P.Bak, C.Tak, K.  Wiesenfeld, Phys.  Rev.  Lett.  {\bf
59}, 381 (1987); Phys. Rev. A {\bf 38}, 364 (1988).

\end{references}
\end{document}